\documentclass[12pt]{article}
\usepackage{amssymb,amsmath,epsfig}

\begin{document}

\title{\bf Energy Content of Colliding Plane Waves using Approximate
Noether Symmetries}

\author{M. Sharif \thanks{msharif.math@pu.edu.pk} and Saira Waheed
\thanks{sairawaheed\_50@yahoo.com}\\
Department of Mathematics, University of the Punjab,\\
Quaid-e-Azam Campus, Lahore-54590, Pakistan.}

\date{}

\maketitle
\begin{abstract}
This paper is devoted to study the energy content of colliding plane
waves using approximate Noether symmetries. For this purpose, we use
approximate Lie symmetry method of Lagrangian for differential
equations. We formulate the first-order perturbed Lagrangian for
colliding plane electromagnetic and gravitational waves. It is shown
that in both cases, there does not exist non-trivial first-order
approximate symmetry generator.
\end{abstract}

{\bf Keywords:} Colliding plane waves; Approximate symmetries;
Conserved quantities.

\section{Introduction}

For the description of a physical system, the conserved quantities
like energy, linear and angular momentum etc. are of great
importance. Being a non-conserved quantity, energy and its
localization is the most prominent and fundamental issue in general
relativity. For gravitational waves (the ripples produced by
accelerating mass in the fabric of the spacetime which propagates
through the space at the speed of light \cite{1}), this problem is
of particular interest. The gravitational waves, by definition, have
zero stress-energy tensor. Thus the existence of these waves was
questioned. However, the theory of general relativity indicates the
existence of gravitational waves as solutions of the Einstein's
field equations \cite{2}. In fact this problem arises because energy
is not well-defined in general relativity. The problem for plane
gravitational waves was resolved by Ehlers and Kundt \cite{3},
Pirani \cite{4} and Weber and Wheeler \cite{4a} by considering a
sphere of test particles in the path of the waves. They showed that
these particles acquired a constant momentum from the waves. Qadir
and Sharif \cite{5} presented an operational procedure, embodying
the same principle, to show that gravitational waves impart a
momentum.

To find a satisfactory solution of this challenging problem,
enormous efforts have been made. The pseudo-tensors which are
firstly introduced by Einstein provide a mean to determine global
energy-momentum conservation. This attracted many others physicists
to develop different prescriptions e.g. \cite{6}-\cite{12}. Being a
combination of $T^{b}_{a}$ and a pseudo-tensor $t^{b}_{a}$ (as it
explicates the energy and momentum density of the gravitational
field), none of these non-tensorial complexes are proved to be
unambiguous.

Recently, an alternative approach, i.e., concept of approximate
symmetry, is used to define energy of gravitational waves. There
have been a number of definitions of approximate symmetry and
different methods \cite{13,14} are available to find these
symmetries. Using this concept, a measure of the extent of the
break-down of time-translational symmetry was presented
\cite{15}-\cite{17}. This yields the so-called an almost symmetric
space and almost Killing vector (the vector field corresponding to
almost symmetric space) \cite{18}. However, each of these attempts
have their own drawbacks.

Noether symmetries also known as symmetries of the Lagrangian,
provide a systematic treatment for the solution of DE arising in
many practical problems \cite{19}. Noether and Lie symmetries have
many important applications like the linearization of the non-linear
equation, the reduction of the order of ordinary differential
equations (ODEs) as well as the number of independent variables of
partial differential equations (PDEs) etc. The double reduction of
DEs and the existence of conservation laws or first integrals
(Noether invariants) are other interesting features of the Noether
symmetries \cite{20}. The more symmetries an equation possesses, the
more easier will be its integration. The field equations that
provide a basis for general theory of relativity (GR) are highly
non-linear coupled PDEs. Finding an exact solution to these field
equations is one of the major problems in GR. The Noether symmetries
approach has been proved to be fruitful in this regard due to their
wide range of applications in the field of DEs \cite{21,22}. It is
worth mentioning here that most of physically interesting solutions
to the field equations, available in literature, possess some kind
of symmetry \cite{23}.

The Killing vectors (KVs) form a subalgebra of Noether symmetries
which further form a subalgebra of Lie point symmetries. By
investigating Noether symmetries and the corresponding conservation
laws for some static models, it was shown that Noether symmetries
yield some non-trivial conservation laws that are different from KVs
\cite{21}. The perturbed Lagrangian yields the approximate
symmetries and the approximate conserved quantities as constants of
motion. By contracting the first order non-trivial approximate
symmetry with the momentum 4-vector, one can obtain energy
non-conservation due to time variation \cite{24}. Since Noether
theorem leads to the fact that each continuous symmetry generator of
the Lagrangian corresponds to a conserved quantity. Therefore, in
order to define energy, it is very useful to utilize the concept of
time symmetry (due to the fact that energy conservation comes from
the time translational invariance).

In going from Minkowski spacetime (flat) to non-flat spacetimes,
some of the conservation laws are lost. Using the connection
between spacetime isometries and its related differential
equations (geodesics) \cite{25,26} for different spacetimes, the
lost conservation laws are recovered and the corresponding energy
re-scaling factors are obtained \cite{27}-\cite{30}. We have used
this procedure to discuss energy in the stringy charged black hole
solutions \cite{31a} and Bardeen model \cite{31b}. In a recent
paper \cite{24}, energy contents of pp and cylindrical
gravitational waves have been discussed by using the same
procedure. In this paper, we extend this procedure to colliding
plane electromagnetic and gravitational waves to check the
existence of conserved quantity.

The plan of this paper is as follows. In the next section, we review
the mathematical framework for exact and approximate symmetry
methods for the Lagrangian of differential equations (DEs). Sections
\textbf{3} and \textbf{4} contain approximate Noether symmetries of
colliding plane electromagnetic and gravitational waves
respectively. The last section provides summary and results.

\section{Mathematical Formulation}

Noether's theorem was proved in 1915 and published in 1918 by Emmy
Noether. This theorem describes the connections between continuous
symmetries of a physical system and their corresponding conserved
quantities \cite{32}. By the analysis of various transformations,
Noether's theorem provides amazing insights into any general theory.
This theorem provides information regarding the conservation laws.
If the results of a physical experiment are independent of any
change in position (homogeneity in space) and time, then the
formulated Lagrangian is symmetric under continuous spacetime
translations and this theorem leads to laws of energy and linear
momentum conservation \cite{33,34}. Similarly, if the physical
experiment is independent of any change in the measurement of angle,
then the physical system is rotationally symmetric and this theorem
implies the conservation law of angular momentum.

Noether's theorem has a wide range of applications in theoretical
physics and calculus of variation. Dissipative system is one of the
examples of the systems which cannot be modeled with Lagrangian.
Therefore, Noether's theorem is inapplicable to these systems and
consequently their continuous symmetries provide no conservation
laws. The original form of this theorem is valid for those
Lagrangian which contain only the first-order derivatives. In many
practical problems, DEs arise with a small term called the
perturbation parameter which indicates a small error or correction.
The purpose for introducing this small term is to examine the given
DE in some certain limit. The Lie point symmetries of such DEs
(perturbed) are of great importance. The application of Noether
symmetry analysis to Lagrangian of a system provides those
symmetries which directly yield our desired conserved quantities. In
that case, one needs to use only the first-order prolongation of the
symmetry generator. The procedure for calculating the symmetries of
the Lagrangian \cite{35,36} is given as follows.

We take a vector field \textbf{X} as
\begin{equation*}
\textbf{X}=\xi(s,x^{a})\frac{\partial}{\partial
s}+\eta^{b}(s,x^{a})\frac{\partial}{\partial x^{b}}
\end{equation*}
and its first-order prolongation is defined as
\begin{equation}\label{1}
\textbf{X}^{[1]}=\textbf{X}+(\eta^{b}_{,s}+\eta^{b}_{,a}\dot{x}^{a}-\xi_{,s}\dot{x}^{b}
-\xi_{,a}\dot{x}^{a}\dot{x}^{b})\frac{\partial}{\partial\dot{x}^{b}},\quad
(a,b=0,1,2,3).
\end{equation}
Suppose a second-order ODE, i.e., Euler-Lagrange equation is given
by
\begin{equation}\label{2}
\ddot{x}^{a}=g(s,x^{a},\dot{x}^{a}).
\end{equation}
The vector field $\textbf{X}$ is said to be Noether point symmetry
of the Lagrangian $L(s,x^{a},\dot{x}^{a})$ (corresponding to
Eq.(\ref{2})), if there exists a function $A(s,x^{a})$ such that the
following condition is satisfied
\begin{equation}\label{3}
\textbf{X}^{[1]}L+(D_{s}\xi)L=D_{s}A.
\end{equation}
Here $D_{s}$ is the total derivative operator given by
\begin{equation}\label{4}
D_{s}=\frac{\partial}{\partial
s}+\dot{x}^{a}\frac{\partial}{\partial x^{a}}
\end{equation}
and $A$ is a gauge function.

The significance of Noether symmetries is given by the following
theorem.\\\\
\textbf{Theorem:} Suppose $L(s,x^{a},\dot{x}^{a})$ be the Lagrangian
corresponding to second-order ODE given by Eq.(\ref{2}) and
$\textbf{X}$ is its corresponding Noether point symmetry. Then the
first integral of motion associated with the Noether point symmetry
$\textbf{X}$ is defined \cite{37} as
\begin{equation}\label{5} I=\xi
L+(\eta^{a}-\dot{x}^{a}\xi)\frac{\partial L}{\partial
\dot{x}^{a}}-A.
\end{equation}
Now we define approximate Noether symmetries for DEs. The
first-order perturbed Lagrangian corresponding to the first-order
perturbed DE
\begin{equation*}
\textbf{E}=\textbf{E}_{0}+\epsilon \textbf{E}_{1}+O(\epsilon^{2})
\end{equation*}
is given by
\begin{equation}\label{5}
L(s,~x^{a},\dot{x}^{a},\epsilon)=L_{0}(s,x^{a},\dot{x}^{a})+\epsilon
L_{1}(s,x^{a},\dot{x}^{a})+O(\epsilon^{2})
\end{equation}
such that the functional $\int_{V}Lds$ is invariant under the
one-parameter group of transformations with approximate Lie
symmetry generator
\begin{equation*}
\textbf{X}=\textbf{X}_{0}+\epsilon \textbf{X}_{1}+ O(\epsilon^{2})
\end{equation*}
and
\begin{equation*}
A=A_{0}+\epsilon A_{1}
\end{equation*}
is the first-order perturbed gauge function. Also, here
\begin{equation*}
\textbf{X}_{b}=\xi_{b}\frac{\partial}{\partial
s}+\eta^{a}_{b}\frac{\partial}{\partial x^{a}};\quad a=0,1,2,3,~
b=0,1.
\end{equation*}
Then the conditions for calculating exact and first-order
symmetries are given as follows
\begin{eqnarray}\label{6}
&&\textbf{X}^{[1]}_{0}L_{0}+(D_{s}\xi_{0})L_{0}=D_{s}A_{0},\\\label{7}
&&\textbf{X}^{[1]}_{1}L_{0}+\textbf{X}^{[1]}_{0}L_{1}
+(D_{s}\xi_{1})L_{0}+(D_{s}\xi_{0})L_{1}=D_{s}A_{1},
\end{eqnarray}
where $\textbf{X}_{0}$ and $\textbf{X}_{1}$ are the exact and
first-order parts, respectively, of the symmetry generator
$\textbf{X}$. The perturbed Lagrangian always admits a symmetry
$\epsilon \textbf{X}_{0}$ called the trivial symmetry. If, for the
symmetry generator $\textbf{X}$ to exist with $\textbf{X}_{0}\neq
0$ and $\textbf{X}_{1}\neq k\textbf{X}_{0}$, where $k$ is any
arbitrary constant, then $\textbf{X}$ is called the non-trivial
symmetry. The corresponding exact and the first-order approximate
parts of the first integral are given by
\begin{eqnarray}\label{8}
I_{0}&=&\xi_{0}
L_{0}+(\eta^{a}_{0}-\dot{x}^{a}\xi_{0})\frac{\partial
L_{0}}{\partial \dot{x}^{a}}-A_{0},\\\label{9}
I_{1}&=&\xi_{0}L_{1}+\xi_{1}L_{0}+(\eta^{a}_{0}-\dot{x}^{a}\xi_{0})\frac{\partial
L_{1}}{\partial\dot{x}^{a}}+(\eta^{a}_{1}-\dot{x}^{a}\xi_{1})\frac{\partial
L_{0}}{\partial \dot{x}^{a}}-A_{1}.
\end{eqnarray}
The first integral of motion $I=I_{0}+\epsilon I_{1}$ is said to
be unstable if $I_{0}=0$; otherwise it is stable.

For further details, one can see the literature
\cite{37}-\cite{39}.

\subsection{Exact and Approximate Symmetries of
the Colliding Plane Electromagnetic Waves}

Bell and Szekeres, in $1974$, presented an exact solution of field
equations due to the collision and the subsequent interaction of
two electromagnetic waves \cite{40}. They considered the case of
two-step electromagnetic waves which collide on the flat
background region. The spacetime is defined in terms of four
regions \cite{41}. Under the coordinate transformations
$u=\frac{t-z}{2},~ v=\frac{t+z}{2}$ \cite{42}, the spacetimes in
these four regions can be written as
\begin{eqnarray}\nonumber
&&\textbf{Region(I)}~(t<z,~t<-z):\\\label{10}
&&ds^{2}=\frac{dt^{2}}{2}-dx^{2}-dy^{2}-\frac{dz^{2}}{2},\\\nonumber
&&\textbf{Region(II)}~(t>z,~t<-z):\\\label{11}
&&ds^{2}=\frac{dt^{2}}{2}-\cos^{2}a(\frac{t-z}{2})(dx^{2}+dy^{2})-\frac{dz^{2}}{2},\\\nonumber
&&\textbf{Region(III)}~(t<z,~t>-z):\\\label{12}
&&ds^{2}=\frac{dt^{2}}{2}-\cos^{2}b(\frac{t+z}{2})(dx^{2}+dy^{2})-\frac{dz^{2}}{2},\\\nonumber
&&\textbf{Region(IV)}~(t>z,~t>-z):\\\label{13}
&&ds^{2}=\frac{dt^{2}}{2}-\cos^{2}(At-Bz)dx^{2}-\cos^{2}(Az-Bt)dy^{2}-\frac{dz^{2}}{2},
\end{eqnarray}
where $A=\frac{a-b}{2}$ and $B=\frac{a+b}{2}$. The solution has a
special property of being conformally flat, i.e., all components
of the Weyl tensor vanish.

In order to discuss Noether symmetries for colliding plane
electromagnetic waves, we define Lagrangian for the spacetime given
by Eq.(\ref{13}) as follows
\begin{equation}\label{14}
L=\frac{\dot{t}^{2}}{2}-\cos^{2}(At-Bz)\dot{x}^{2}-\cos^{2}(Az-Bt)\dot{y}^{2}
-\frac{\dot{z}^{2}}{2}.
\end{equation}
To evaluate exact Noether symmetries of this Lagrangian, we
substitute values of the Lagrangian, the first-order prolongation of
the symmetry generator (given by Eq.(\ref{1})) and the total
derivative operator (given by Eq.(\ref{4})) in Eq.(\ref{6}). Then we
compare coefficients of the coordinate derivatives and their
products on both sides. In this way, we obtain the following set of
DEs
\begin{eqnarray}\label{15}
&&\xi_{,t}=0,\quad \xi_{,x}=0,\quad \xi_{,y}=0,\quad
\xi_{,z}=0,\quad A_{,s}=0,
\\\label{16}
&&\eta^{0}_{,s}=A_{,t},\quad
-2\cos^{2}(At-Bz)\eta^{1}_{,s}=A_{,x},\\\label{17}
&&-2\cos^{2}(Az-Bt)\eta^{2}_{,s}=A_{,y},\quad-\eta^{3}_{,s}=A_{,z},\\\label{18}
&&\eta^{0}_{,x}-2\cos^{2}(At-Bz)\eta^{1}_{,t}=0,\\\label{19}
&&\eta^{0}_{,y}-2\cos^{2}(Az-Bt)\eta^{2}_{,t}=0,\quad
\eta^{0}_{,z}-\eta^{3}_{,t}=0, \\\label{20}
&&\cos^{2}(At-Bz)\eta^{1}_{,y}+\cos^{2}(Az-Bt)\eta^{2}_{,x}=0,\\\label{21}
&&\eta^{3}_{,x}+2\cos^{2}(At-Bz)\eta^{1}_{,z}=0,\quad2\eta^{0}_{,t}-\xi_{,s}=0,\\\label{22}
&&\eta^{3}_{,y}+2\cos^{2}(Az-Bt)\eta^{2}_{,z}=0,\quad
2\eta^{3}_{,z}-\xi_{,s}=0,\\\label{23}
&&2\eta^{1}_{,x}-\xi_{,s}-2(A-B)\tan(At-Bz)(\eta^{0}+\eta^{3})=0,\\\label{24}
&&2\eta^{2}_{,y}-\xi_{,s}-2(A-B)\tan(Az-Bt)(\eta^{0}+\eta^{3}) =0.
\end{eqnarray}
This is a system of $19$ DEs which we have to solve for the five
unknowns $\xi,~\eta^{0},~\eta^{1},~\eta^{2},~\eta^{3}$ using back
and forth substitutions (where each of these is a function of five
variables $s,~t,~x,~y,~z$). The solution (exact symmetry generators)
to the above set of DEs can be written as
\begin{equation}\label{25}
\textbf{Y}_{0}=s\frac{\partial}{\partial
t}-s\frac{\partial}{\partial z},\quad
\textbf{Y}_{1}=\frac{\partial}{\partial x},\quad
\textbf{Y}_{2}=\frac{\partial}{\partial y}.
\end{equation}

It has been mentioned earlier that the electromagnetic waves collide
with each other on the flat background region (i.e., Minkowski
spacetime in Cartesian coordinates) and is represented by non-static
spacetime (given by Eq.(\ref{13})). Therefore we define it as a
perturbation over some static spacetime, i.e.,  spacetime
representing the interaction region (region (\textbf{IV})) as a
perturbation over the flat Minkowski spacetime. The exact static
spacetime is given by
\begin{equation}\label{26}
ds^{2}=\frac{dt^{2}}{2}-dx^{2}-dy^{2}-\frac{dz^{2}}{2}
\end{equation}
and the corresponding Lagrangian can be written as
\begin{equation}\label{27}
L=\frac{\dot{t}^{2}}{2}-\dot{x}^{2}-\dot{y}^{2}-\frac{\dot{z}^{2}}{2}.
\end{equation}

We evaluate Noether symmetries of this Lagrangian by using
Eq.(\ref{6}). After substituting all the corresponding values, we
compare coefficients of the coordinate derivatives and their
products and obtain the following set of determining equations
\begin{eqnarray}\label{28}
&&\xi_{,t}=0,\quad \xi_{,x}=0,\quad \xi_{,y}=0,\quad
\xi_{,z}=0,\quad A_{,s}=0,
\\\label{29}
&&\eta^{0}_{,s}=A_{,t},\quad-2\eta^{1}_{,s}=A_{,x},\quad-\eta^{3}_{,s}=A_{,z},\\\label{30}
&&-2\eta^{2}_{,s}=A_{,y},\quad
\eta^{0}_{,x}-2\eta^{1}_{,t}=0,\\\label{31}
&&\eta^{0}_{,y}-2\eta^{2}_{,t}=0,\quad
\eta^{0}_{,z}-\eta^{3}_{,t}=0,\quad
\eta^{1}_{,y}+\eta^{2}_{,x}=0,\\\label{32}
&&\eta^{3}_{,x}+2\eta^{1}_{,z}=0,\quad2\eta^{0}_{,t}-\xi_{,s}=0,\\\label{33}
&&\eta^{3}_{,y}+2\eta^{2}_{,z}=0,\quad 2\eta^{3}_{,z}-\xi_{,s}=0,
\\\label{34}
&&2\eta^{1}_{,x}-\xi_{,s}=0,\quad 2\eta^{2}_{,y}-\xi_{,s}=0.
\end{eqnarray}
The simultaneous solution for this set of DEs is given by
\begin{eqnarray}\label{35}
\xi&=&\frac{c_{0}s^{2}}{2}+c_{1}s+c_{2},\\\label{36}
\eta^{0}&=&\frac{t(c_{0}s+c_{1})}{2}+sc_{3}+2xc_{5}+2c_{7}y+c_{12}z+c_{13},\\\label{37}
\eta^{1}&=&\frac{x(c_{0}s+c_{1})}{2}+sc_{4}+tc_{5}+c_{8}y+c_{14}z+c_{15},\\\label{38}
\eta^{2}&=&\frac{y(c_{0}s+c_{1})}{2}+sc_{6}+tc_{7}+c_{16}z-xc_{8}+c_{17},
\\\label{39}
\eta^{3}&=&\frac{z(c_{0}s+c_{1})}{2}-sc_{9}+tc_{12}-2xc_{14}-2yc_{16}+c_{11},\\\label{40}
A(t,x,y,z)&=&c_{0}(\frac{t^{2}}{4}-\frac{x^{2}}{2}-\frac{y^{2}}{2}-\frac{z^{2}}{4})
+tc_{3}-2xc_{4}-2yc_{6}+c_{9}z+c_{10},\nonumber\\
\end{eqnarray}
where all c's are arbitrary constants of integration. Clearly this
is a $17$ dimensional algebra in which many symmetries are present
which provide no conservation laws. In this algebra, $10$
symmetries correspond to the generators which form the Poincare
group and $7$ provide the other significant generators which have
been discussed in detail \cite{29}. These symmetries can be
re-arranged to the following form
\begin{eqnarray}\nonumber
\textbf{Y}_{0}&=&\frac{st}{2}\frac{\partial}{\partial
t}+\frac{sx}{2}\frac{\partial}{\partial
x}+\frac{sy}{2}\frac{\partial}{\partial
y}+\frac{sz}{2}\frac{\partial}{\partial z},\\\nonumber
\textbf{Y}_{1}&=&\frac{Y_{0}}{s},~
\textbf{Y}_{2}=s\frac{\partial}{\partial
t},~\textbf{Y}_{3}=2x\frac{\partial}{\partial
t}+t\frac{\partial}{\partial x},\\\nonumber
\textbf{Y}_{4}&=&s\frac{\partial}{\partial
y},~\textbf{Y}_{5}=\frac{\partial}{\partial
x},~\textbf{Y}_{6}=2y\frac{\partial}{\partial
t}+t\frac{\partial}{\partial y},\\\nonumber
\textbf{Y}_{7}&=&y\frac{\partial}{\partial
x}-x\frac{\partial}{\partial
y},~\textbf{Y}_{8}=-s\frac{\partial}{\partial z},~
\textbf{Y}_{9}=\frac{\partial}{\partial z},\\\nonumber
\textbf{Y}_{10}&=&z\frac{\partial}{\partial
t}+t\frac{\partial}{\partial
z},~\textbf{Y}_{11}=\frac{\partial}{\partial t},\\\nonumber
\textbf{Y}_{12}&=&z\frac{\partial}{\partial
x}-2x\frac{\partial}{\partial
z},~\textbf{Y}_{13}=\frac{\partial}{\partial x},\\\nonumber
\textbf{Y}_{14}&=&z\frac{\partial}{\partial
y}-2y\frac{\partial}{\partial z},~
\textbf{Y}_{15}=\frac{\partial}{\partial y},
\textbf{Y}_{16}=s\frac{\partial}{\partial x}.
\end{eqnarray}

In order to calculate conserved quantity by using the non-trivial
first-order Noether symmetry (if it exists), we define the perturbed
spacetime as follows
\begin{equation}\label{51}
ds^{2}=\frac{dt^{2}}{2}-[1+\epsilon\cos^{2}(At-Bz)]dx^{2}
-[1+\epsilon\cos^{2}(Az-Bt)]dy^{2}-\frac{dz^{2}}{2}.
\end{equation}
(Basically it means that we are taking the spacetime of region
(\textbf{IV}) as a perturbation over the spacetime of region
(\textbf{I})). The corresponding first-order perturbed Lagrangian is
given by
\begin{equation}\label{52}
L=\frac{\dot{t}^{2}}{2}-[1+\epsilon\cos^{2}(At-Bz)]\dot{x}^{2}
-[1+\epsilon\cos^{2}(Az-Bt)]\dot{y}^{2}-\frac{\dot{z}^{2}}{2}.
\end{equation}
The first-order approximate symmetries of this Lagrangian is found
by using Eq.(\ref{7}). After substituting the Lagrangian and other
corresponding values, we make comparison of the coefficients of
coordinate derivatives and their product terms. Consequently we
obtain the following set of DEs
\begin{eqnarray}\label{54}
&&\xi_{,t}=0,\quad \xi_{,x}=0,\quad \xi_{,y}=0,\quad
\xi_{,z}=0,\quad A_{,s}=0,
\\\label{55}
&&\eta^{0}_{,s}=A_{,t},\quad
-2\eta^{1}_{,s}-2\cos^{2}(At-Bz)[\frac{xc_{0}}{2}+c_{4}]=A_{,x},\\\label{56}
&&-2\eta^{2}_{,s}-\cos^{2}(Az-Bt)[\frac{yc_{0}}{2}+c_{6}]=A_{,y},\quad-\eta^{3}_{,s}=A_{,z},\\\label{57}
&&\eta^{0}_{,x}-2\eta^{1}_{,t}-2c_{5}\cos^{2}(At-Bz)=0,\\\label{58}
&&\eta^{0}_{,y}-2\eta^{2}_{,t}-2c_{7}\cos^{2}(Az-Bt)=0,\quad
\eta^{0}_{,z}-\eta^{3}_{,t}=0,\\\label{59}
&&\eta^{1}_{,y}+\eta^{2}_{,x}+c_{8}[\cos^{2}(At-Bz)-\cos^{2}(Az-Bt)]=0,\\\label{60}
&&\eta^{3}_{,x}+2\eta^{1}_{,z}+2c_{14}\cos^{2}(At-Bz)=0,
~2\eta^{0}_{,t}-\xi_{,s}=0,\\\label{61}
&&\eta^{3}_{,y}+2\eta^{2}_{,z}+2c_{16}\cos^{2}(Az-Bt)=0,\quad
2\eta^{3}_{,z}-\xi_{,s}=0,\\\nonumber
&&2\eta^{1}_{,x}-\xi_{,s}-2(A-B)\cos(At-Bz)\sin(At-Bz)[\frac{t(c_{0}s+
c_{1})}{2}\\\nonumber\ &&+\frac{z(c_{0}s+c_{1})}{2}
+s(c_{3}-c_{9})+2x(c_{5}-c_{14})+2y(c_{7}-c_{16})\\\label{62}
&&+c_{12}(t+z)
+c_{13}+c_{11}]+(c_{0}s+c_{1})\cos^{2}(At-Bz)=0,\\\nonumber
&&2\eta^{2}_{,y}-\xi_{,s}-2(A-B)\cos(Az-Bt)\sin(Az-Bt)[\frac{t(c_{0}s+c_{1})}{2}\\\nonumber
&&+\frac{z(c_{0}s+c_{1})}{2}+s(c_{3}-c_{9})+2x(c_{5}-c_{14})+2y(c_{7}-c_{16})\\\label{63}
&&+c_{12}(t+z)+c_{13}+c_{11}]-(c_{0}s+c_{1})\cos^{2}(Az-Bt)=0 =0.
\end{eqnarray}
In this system of DEs, $14$ out of $17$ constants corresponding to
exact symmetry generators are present. We solve this system of DEs
by back and forth substitutions and check whether there exist some
non-trivial parts of the symmetry generators or not. Since all the
constants corresponding to the exact symmetry generators given by
Eqs.(\ref{35})-(\ref{39}) disappear, therefore the system of DEs
given in Eqs.(\ref{54})-(\ref{63}) becomes homogeneous. Thus it
yields symmetries of the static spacetime only.

\subsection{Exact and Approximate Symmetries of the Colliding Plane Gravitational Waves}

The spacetime representing the collision of two plane
gravitational waves in Rosen form is given by
\begin{equation*}
ds^{2}=2e^{-M}dudv-e^{-U}(e^{V}dx^{2}+e^{-V}dy^{2}),
\end{equation*}
where $U,~V$ and $M$ are functions of the null coordinates $u$ and
$v$. For the collision of such waves, one has the following
division of the spacetime into four regions. The spacetimes in
these four regions are given by \cite{43}:
\begin{eqnarray}\nonumber
&&\textbf{Region(I)}~(u<0,~v<0):\\\label{64}
&&ds^{2}=2dudv-dx^{2}-dy^{2},\\\nonumber
&&\textbf{Region(II)}~(u>0,~v<0):\\\label{65} &&ds^{2}=2(1\pm
u)^{\frac{(a^{2}-1)}{2}}dudv-(1\pm u)^{1-a}dx^{2}-(1\pm
u)^{1+a}dy^{2},\\\nonumber
&&\textbf{Region(III)}~(u<0,~v>0):\\\label{66} &&ds^{2}=2(1\pm
v)^{\frac{(a^{2}-1)}{2}}dudv-(1\pm v)^{1-a}dx^{2}-(1\pm
v)^{1+a}dy^{2},\\\nonumber
&&\textbf{Region(IV)}~(u>0,~v>0):\\\nonumber &&ds^{2}=2(1\pm u\pm
v)^{\frac{(a^{2}-1)}{2}}dudv-(1\pm u \pm
v)^{1-a}dx^{2}\\\label{67}&&-(1\pm u\pm v)^{1+a}dy^{2},
\end{eqnarray}
where $a$ is any arbitrary parameter. If one chooses $a=0$ and the
negative signs only, then this solution is quite similar to the
solution obtained by Szekeres and Khan and Penrose \cite{44}.
Using the suitable coordinate transformations
$u=\frac{t-z}{2},~v=\frac{t+z}{2}$ \cite{42} and taking only the
negative signs, we have the spacetime in region \textbf{(IV)} as
follows
\begin{equation}\label{68}
ds^{2}=2(1-t)^{\frac{-1}{2}}(\frac{dt^{2}}{4}-\frac{dz^{2}}{4})
-(1-t)(dx^{2}+dy^{2}).
\end{equation}

To evaluate exact symmetries of the Lagrangian for this spacetime
\begin{equation}\label{69}
L=2(1-t)^{\frac{-1}{2}}(\frac{\dot{t^{2}}}{4}-\frac{\dot{z^{2}}}{4})
-(1-t)(\dot{x^{2}}+\dot{y^{2}})
\end{equation}
we use Eq.(\ref{6}) and substitute the values of first-order
prolongation operator, the total derivative operator and the above
defined Lagrangian in it. Then we compare the coefficients of the
derivatives and their products, we obtain the following set of DEs
\begin{eqnarray}\label{70}
&&\xi_{,t}=0,\quad \xi_{,x}=0,\quad \xi_{,y}=0,\quad
\xi_{,z}=0,\quad A_{,s}=0,\\\label{71}
&&\eta^{0}_{,s}\frac{1}{\sqrt{1-t}}=A_{,t},\quad-2(1-t)\eta^{1}_{,s}=A_{,x},\\\label{72}
&&-2(1-t)\eta^{2}_{,s}=A_{,y},\quad-\eta^{3}_{,s}\frac{1}{\sqrt{1-t}}=A_{,z},\\\label{73}
&&\eta^{0}_{,x}-2(1-t)^{3/2}\eta^{1}_{,t}=0,\\\label{74}
&&\eta^{0}_{,y}-2(1-t)^{3/2}\eta^{2}_{,t}=0,\quad
\eta^{0}_{,z}-\eta^{3}_{,t}=0,\\\label{}
&&\eta^{1}_{,y}+\eta^{2}_{,x}=0,\quad\eta^{3}_{,x}+2(1-t)^{3/2}\eta^{1}_{,z}=0,\\\label{75}
&&2(1-t)[2\eta^{0}_{,t}-\xi_{,s}]+\eta^{0}=0,\quad
\eta^{3}_{,y}+2(1-t)^{3/2}\eta^{2}_{,z}=0,\\\label{76}
&&2(1-t)(2\eta^{3}_{,z}-\xi_{,s})-\eta^{0}=0,\\\label{77}
&&(1-t)[2\eta^{1}_{,x}-\xi_{,s}]-\eta^{0}=0,\quad
(1-t)[2\eta^{2}_{,y}-\xi_{,s}]-\eta^{0}=0.
\end{eqnarray}
The simultaneous solution of this system of DEs turns out to be
\begin{eqnarray}\label{78}
\eta^{0}=0,\quad
\eta^{1}=b_{6}y+b_{7},\quad\eta^{2}=-b_{6}x+b_{8},\quad
\eta^{3}=b_{5},\quad\xi_{2}=b_{2},\quad A=b_{4},
\end{eqnarray}
where all $b$'s are constants, which show that the gauge function is
constant. These can be re-arranged to the following form
\begin{eqnarray}\nonumber
A=b_{0},\quad\xi_{2}=c_{0},\quad
\textbf{Y}_{1}=\frac{\partial}{\partial z},\quad
\textbf{Y}_{2}=y\frac{\partial}{\partial
x}-x\frac{\partial}{\partial y},\quad
\textbf{Y}_{3}=\frac{\partial}{\partial x},\quad
\textbf{Y}_{4}=\frac{\partial}{\partial y}.
\end{eqnarray}

In order to discuss the corresponding conserved quantity, we take
Eq.(\ref{68}) as a perturbation over background static spacetime in
a similar pattern as we have done for colliding electromagnetic
waves. The background static spacetime is the Minkowski spacetime in
Cartesian coordinates given by Eq.(\ref{26}) and its Noether
symmetries are given by Eqs.(\ref{35})-(\ref{39}). T evaluate the
approximate Noether symmetries, the first-order perturbed spacetime
for the collision of gravitational waves is given by
\begin{equation*}
ds^{2}=[1+\epsilon
(1-t)^{\frac{-1}{2}}](\frac{dt^{2}}{2}-\frac{dz^{2}}{2})-[1+\epsilon(1-t)](dx^{2}+dy^{2}).
\end{equation*}
The corresponding Lagrangian is formulated as
\begin{equation}\label{79}
L=[1+\epsilon
(1-t)^{\frac{-1}{2}}](\frac{\dot{t^{2}}}{2}-\frac{\dot{z^{2}}}{2})
-[1+\epsilon(1-t)](\dot{x^{2}}+\dot{y^{2}}).
\end{equation}
Using this Lagrangian and the corresponding values in Eq.(\ref{7}),
we obtain the following system of DEs
\begin{eqnarray}\label{80}
&&\xi_{,t}=0,~ \xi_{,x}=0,~ \xi_{,y}=0,~ \xi_{,z}=0,~ A_{,s}=0,
\\\label{}
&&\eta^{0}_{,s}+(\frac{c_{0}t}{2}+c_{3})\frac{1}{\sqrt{1-t}}=A_{,t},\\\label{81}
&&-2\eta^{1}_{,s}-2(1-t)(\frac{xc_{0}}{2}+c_{4})=A_{,x},\\\label{82}
&&-2\eta^{2}_{,s}-2(1-t)(\frac{c_{0}y}{2}+c_{6})=A_{,y},\\\label{83}
&&-\eta^{3}_{,s}-\frac{1}{\sqrt{1-t}}(\frac{c_{0}z}{2}-c_{9})=A_{,z},\\\label{84}
&&[\eta^{0}_{,x}-2\eta^{1}_{,t}]+\frac{2}{\sqrt{1-t}}c_{5}(1-(1-t)^{3/2})=0,\\\label{85}
&&[\eta^{0}_{,y}-2\eta^{2}_{,t}]+\frac{2}{(1-t)^{1/2}}c_{7}(1-(1-t)^{3/2})=0,\\\label{86}
&&\eta^{0}_{,z}-\eta^{3}_{,t}=0,\eta^{1}_{,y}+\eta^{2}_{,x}=0,\\\label{87}
&&[\eta^{3}_{,x}+2\eta^{1}_{,z}]-\frac{2c_{14}}{\sqrt{1-t}}(1-(1-t)^{3/2})=0,\\\nonumber
&&[2\eta^{0}_{,t}-\xi_{,s}]+\frac{1}{2(1-t)^{3/2}}[\frac{t}{2}(c_{0}s+c_{1})
+sc_{3}+2xc_{5}+2c_{7}y\\\label{88}
&&+c_{12}z+c_{13}]=0,\\\label{89}
&&\eta^{3}_{,y}+2\eta^{2}_{,z}-\frac{2c_{16}}{\sqrt{1-t}}(1-(1-t)^{3/2})=0,\\\nonumber
&&(2\eta^{3}_{,z}-\xi_{,s})+\frac{1}{2(1-t)^{3/2}}[\frac{t}{2}(c_{0}s+c_{1})+sc_{3}
+2xc_{5}+2c_{7}y\\\label{90}&&+c_{12}z +c_{13}]=0,\\\nonumber
&&[2\eta^{1}_{,x}-\xi_{,s}]-(\frac{t}{2}(c_{0}s+c_{1})+sc_{3}+2xc_{5}+2c_{7}y\\\label{91}
&&+c_{12}z+c_{13})=0,\\\nonumber
&&[2\eta^{2}_{,y}-\xi_{,s}]-(\frac{t}{2}(c_{0}s+c_{1})+sc_{3}+2xc_{5}+2c_{7}y\\\label{92}
&&+c_{12}z+c_{13})=0.
\end{eqnarray}
In this system of 19 DEs, 12 out of 17 constants, corresponding to
exact symmetry generators defined in Eqs.(\ref{35})-(\ref{39}),
appear. When we solve this system of DEs simultaneously then in the
final solution all these constants disappear. Therefore it makes the
system of equations homogeneous and no non-trivial symmetry
generator exists.

\section{Summary and Discussion}

In this paper, we have discussed energy contents for the colliding
plane waves. For this purpose, we have applied the slightly broken
or approximate Lie symmetry methods for Lagrangian to colliding
plane electromagnetic and gravitational waves. Since the
contraction of non-trivial approximate Noether symmetry
(corresponding to time) with the momentum 4-vector can be used to
define energy imparted to test particles \cite{24}. Therefore, we
have defined the first-order perturbed Lagrangian for both these
solutions and check whether there exists some non-trivial
symmetry.

We have obtained a system of 19 DEs both for the colliding plane
electromagnetic as well as colliding plane gravitational waves. For
the colliding plane electromagnetic waves, 14 out of 17 constants
whereas for colliding plane gravitational waves, 12 out of 17
constants corresponding to exact symmetry generators defined in the
Eqs.(\ref{35})-(\ref{39}) appear. When we solve these systems of DEs
simultaneously, all these constants disappear in both cases. Thus,
the resulting systems of DEs becomes homogeneous and are the same as
that for the exact case (i.e., Minkowski spacetime in Cartesian
coordinates). Consequently only the previous exact symmetry
generators of static spacetime which correspond to 17 dimensional
Lie algebra for Minkowski sapcetime in Cartesian coordinates are
obtained. These symmetry generators provide the conservation laws
for energy, linear momentum, angular and spin angular momentum and
hence there does not exist any non-trivial symmetry.

Thus we conclude that for these first-order perturbed Lagrangian
for colliding plane electromagnetic and gravitational waves, there
does not exist the conserved quantity. It is quite similar to the
case of pp and cylindrical gravitational waves \cite{24} where no
non-trivial symmetry generator exists and one cannot determine the
conserved quantity. It is mentioned here that one cannot apply
these methods to the case of electromagnetic waves as there is no
self-interaction. However, it may give significant result about
the self-damping or enhancement of the gravitational waves if one
applies this method to the system of geodesic equations for these
spacetimes.

\end{document}